# Breakdown of Fast Water Transport in Graphene Oxides

Ning Wei and Zhiping Xu[*]

Applied Mechanics Laboratory, Department of Engineering Mechanics and Center for Nano and Micro Mechanics, Tsinghua University, Beijing 100084, China

[*]Email: xuzp@tsinghua.edu.cn

## ABSTRACT

Fast slip flow was reported for water inside the interlayer gallery between graphene layers or carbon nanotubes. We report here that this flow rate enhancement (over two orders) breaks down with the presence of chemical functionalization and relaxation of the nanoconfinement in graphene oxides. Molecular dynamics simulation results show that hydrodynamics applies in this circumstance, even at length scales down to nanometers. However, corrections on the slip boundary condition and viscosity of nanoconfined flow must be included to make quantitative predictions. These results were discussed with structural characteristics of the liquid water and hydrogen bond networks.

# I. INTRODUCTION

Recent studies have revealed ultrafast fluid flow in hydrophobic nanoconfined environment, e.g. inside carbon nanotubes and interlayer gallery between graphene sheets [1-3]. The remarkably enhanced water transport deviates distinctly from the prediction using macroscopic viscous flow models, and yet holds great promises in filtration and energy conversion applications [4]. The observation was mainly explained by referring to the low friction between water and the graphitic surface where notable interfacial slip occurs [1-3, 5-7]. The flow enhancement was also attributed to the modified water structures of water in the nano-sized channel, which is a hallmark of nanofined liquid [8, 9]. A curvature-induced effect was reported in a comparative study between graphene and carbon nanotubes with different channel widths [3]. 'Structured water' was identified for flow inside carbon nanotubes with diameters less than 1.25 nm [10], including single-file chains, tilted, stacked pentagons, and hexagons. However, all these studies were conducted in nanochannels with atomistically smooth graphitic walls, and the critical effect from the atomistic roughness, such as those created by imperfections and chemical functionalization, was completed excluded [7]. In contrast, an exemplified study recently showed that for water flow between graphene oxide (GO) sheets, oxidized region prohibits fast water transport [11]. Instead, a two-dimensional network could form between the pristine graphene region in GO sheets, still featuring significant flow enhancement, and a capillary-driven flow mechanism was proposed to explain the unusually high water permeance [11].

To clarify the roles of surface functionalization in combination with the nanoconfinement, we performed pressure-driven flow simulations to quantify the nature of interlayer flow between graphene and GO sheets respectively. The structures of liquid water and hydrogen bond (H-bond) networks between them that correlate with both functionalization and nanoconfinement was analyzed and discussed. We find that very low concentration (~5%) of hydroxyl groups on the graphene sheet reduces the slip length by 97%, from 48.13 to 1.26 nm. The dependence of flow rate on the pressure gradient follows the Poiseuille law for conventional viscous flow only after corrections on the viscosity and interfacial slip condition have been made.

## II. MODELS AND METHODS

The molecular structure of graphene oxides consists of hydroxyl, epoxy, and carbonyl groups on the basal plane, defective sites and open edges, where the hydroxyl groups were reported to be able to stay in rich in the long-living quasi-equilibrium state [12]. A typical fraction of hydroxyl species relative to the amount of carbon atoms in GO is ~20% [12]. Further reduction could yield lower concentration (13.9%-15.9%) in the reduced graphene oxide (RGO) [13]. For simplicity, we constructed hydroxyl-functionalized graphene (on both sides of the sheet) with various concentration $c$, defined as $c = n_{OH}/n_C$ ($n_{OH}$ and $n_C$ are the numbers of hydroxyl groups and carbon atoms, respectively). The distribution of hydroxyl groups was sampled randomly.

Empirical molecular dynamics (MD) simulations were performed using the large-scale atomic/molecular massively parallel simulator (LAMMPS) [14]. The optimized all-atom forcefield OPLS-AA was used for GO, which is able to capture essential many-body terms in inter-atomic interactions, i.e. bond stretching, bond angle bending, van der Waals and electrostatic interactions [15]. Following previous studies on similar systems, the rigid SPC/E model for water molecules was used [3, 16]. SHAKE algorithm was applied for the stretching terms between oxygen and hydrogen atoms to reduce high-frequency vibrations that require shorter time steps. The interaction between water and GO includes both van der Waals and electrostatic terms. The former one was described by the Lennard-Jones potential between oxygen and carbon atoms with parameters $\varepsilon$ = 4.063 meV and $\sigma$ = 0.319 nm [17]. Periodic boundary conditions were applied on all directions. The channel length along the flow ($x$) direction is 12.3 nm, and the width is 2.1 nm. In order to ensure that the channel is filled at 1 a.t.m., we adjusted the interlayer distance accordingly at room temperature (300 K) for specific number of atoms. For example, there are 529, 1136, and 2354 water molecules included for interlayer distance $h$ of 1.0, 1.7, and 3.1 nm, respectively. Here $h$ is the distance measured from basal planes in GO. As will be discussed later, the nature of flow strongly correlates with the liquid-solid interfacial properties, one of the key parameters for which is the contact angle $\theta_c$. This set of parameters predicts a contact angle of $95^o$ in consistent with experimental evidence [18]. The van der Waals forces were truncated at 1.0 nm and the long-range Coulomb interactions were computed by using the particle-particle particle-mesh (PPPM) algorithm. To validate our model for GO, we further calculated the contact angle $\theta_c$ between water and graphene sheet with $c$ up to 20%. The results show that $\theta_c$ decrease as $c$ increases (e.g. $\theta_c = 29.1^o$ at $c$ = 20%). This result is consistent with recent experimental measurement [19], and validates the combined OPLS-AA and SPC/E approach here can yield reasonable prediction for the water/GO hybrid system.

The pressure-driven water flow was simulated by directly applying forces to water molecules. This non-equilibrium molecular dynamics (NEMD) approach was widely used to explore fluid flow [16, 20]. All simulations here were performed in the microcanonical ensemble after well-converged equilibration at 300 K using the Berendsen thermostat. The carbon atoms in GO were frozen in the dynamic simulations to maintain the planar conformation of GO sheet as stacked in a paper or thin-film form. After the flow is driven, it usually takes a few nanoseconds to reach the steady flow state, where the external driving force and friction balance. Our following discussion is based on simulation data collected in the subsequent stages.

## III. RESULTS AND DISCUSSION

**A. Interfacial slippage.**

We first explored the interfacial slippage between water the GO walls. To quantify it, the solid-liquid friction coefficient $\lambda$ was calculated from the frictional force autocorrelation function in the equilibrium molecular dynamics (EMD) simulations, following the Green-Kubo formulation

$$\lambda = \frac{1}{Sk_BT}\int_0^\infty \langle F_x(t)F_x(0)\rangle \mathrm{d}t, \qquad (1)$$

where $F_x(t)$ is the time-dependent total force acting on the surface with area $S$, which could be then related to the slip length $l_s$ through $l_s = \eta/\lambda$.[3, 5] A shear viscosity $\eta = 0.729$ mPa.s (for SPC/E water model at 300 K) could used for the confined water in the evaluation [21]. However, it should be noticed that the viscosity could be modified by the nanoconfinement [6, 22, 23]. Thus further corrections from the interlayer distance between GO sheets should be, and was included in our work, for accurate prediction of $\lambda$ (see the **Supplementary Material** for more discussion). The results are plotted in **Figure 1a**, which indicate that the slip length of water flow between pristine graphene sheets is 48.1 nm, and is shortened drastically, by two orders, to 0.44 nm for $c = 30\%$. As the flow enhancement factor $\varepsilon = 1 + 6l_s/d$ for Poiseuille flow with interplate distance $d$, $\varepsilon$ is then reduced by two orders as well. Moreover, we also found that the interfacial slip is enhanced as the contact angle $\theta_c$ increases, following a scaling relation $l_s \sim \tau_F(1 + \cos \theta_c)^{-2}$, where is the relaxation time for confined water (see the **Supplementary Material** and **Figure S1**). This correlation suggests a direct impact from molecular interaction between water and structures, which is much enhanced in the hydrophilic GO and increases with $c$ because of the electrostatic forces between water and hydroxyl groups.

The interfacial slip was also calculated according to the definition of Navier slip length $l_s = v_s / (dv/dz)|_{z=0}$, where $v_s$ is the slip velocity at the fluid-wall interface and $dv/dz$ is the tangent of velocity profile along the normal direction $z$. The exact position of the interface ($z = 0$) is defined as the averaged position of the first water layer from GO sheets. $l_s$ extracted from the velocity profiles in our NEMD simulations is plotted in **Figure 1b** for different molecular structures and driving forces. Our results show that the NEMD results are consistent with the correlation function based ones in equilibrium (**Figure S2**), which suggests the NEMD actually captures the essential fluid dynamics at the interface. This consistency also confirms the significant suppression of water slippage by the presence of hydroxyl groups at a low concentration.

**B. The liquid structure of interlayer water.**

In addition to the interfacial slip, the liquid structure of nanoconfined water also has strong effect on the molecular transport. The subcontinuum transport mechanism in narrow carbon nanotubes was pointed out by McGaughey and other researchers recently [3, 10], who suggested the existence of several low-dimensional, ordered forms of 'structured water' in nanochannels such as the interior space of carbon nanotubes with diameter ~1 nm, e.g. wires and nanotubes [8, 9]. In comparison, the spatial constriction is much released in the 2D gallery space between graphitic layers. However, contrastive structures of the first few water molecular layers from the sheet are expected. From the spatial density of water molecules calculated from MD simulations, single, bilayer, and trilayer structures of water between graphene sheets are clearly identified (**Figure 2a**). For interlayer distance $h$ below 0.7 nm, water monolayer is observed, with H-bond networks forming with hydroxyl groups on GO. Distinct bilayer and trilayer water structures exist below $h$ = 1.0 and 1.4 nm, respectively. With increasing water content, layered order cannot be preserved further and the liquid structure becomes close to the bulk phase ($h$ > 3 nm).

It is interesting to note that although the hydroxyl groups intercalate between water and the carbon basal plane in GO, the shorter interacting range of H-bonds than van der Waals interactions leads to a similar distance from basal plane to the first water layer (L1), whether or not the graphene sheet is functionalized (**Figure 2b**). This is interesting because the water flow between GO with a certain pattern of oxidation groups can then be considered as a continuum without different in the geometry but only modification of the boundary slip condition and structures of water confined therein. To further characterizing the order of H-bond network forming between GO sheets, we analyzed the H-bond network structures and categorizing H-bonds into intralayer and interlayer contributions (the layers are defined with thickness of 0.1 nm). Moreover, in L1 (with average distance ~0.35 nm from the basal plane), water molecules prefer to lie in parallel to the basal plane in the pristine graphene region, corresponding to a peak in the intralayer H-bond distribution (**Figure S3**). For water molecules further from the GO sheet, a more random distribution is identified for their orientations, with the number of intralayer H-bonds per water molecule equals ~1/3 of the peak value for L1. On the other hand, as $c$ increases, more H-bonds could form between water molecules in L1 and the GO sheets, thus the peak in the intralayer H-bond distribution is reduced and the amplitude of distribution in interlayer H-bonds increases. Our simulation results indicate further that the interlayer distance does not modify these structural characteristics. As the nanoconfined water flow between GO sheets critically depends on the first few water layers from GO sheets, which provide the boundary condition for the flow profile, flow characteristics that deviate from bulk water flow would be expected.

**C. The viscous nature of flow.**

With the hydroxyl functionalization on graphene, not only the interfacial slip is suppressed as measured for the reduction of $l_s$, the nature of flow is also modified. To characterize this, we plotted the flow profiles of water in **Figure 3a**. It is clearly shown that, in contrast to the plug-like flow between graphene layers, the velocity profile changes from a plateau ($c < 1\%$) approaching a parabolic one. The trend is even more significant as $c$ increases. The nanofinement reshapes the profile further. Parabolic, continuum viscous flow characteristics are recovered for $c = 10\ \%$ (**Figure 3b**), although there are discrete data points deviating from the ideal behavior for $h$ below 1.0 nm. The breakdown of slip flow by increasing surface functionalization or releasing the nanoconfinement (increasing $h$) is found to be continuous instead of transitional, similar to the changes observed in interfacial slippage and friction.

The parabolicity of the velocity profile implies that the continuum hydrodynamics could apply. The viscous flow between parallel plates can be captured by the Poiseuille solution of the Navier-Stokes equations

$$Q = -(dp/dx)\, d^3/12\eta, \qquad (2)$$

where $Q$ is the flow rate, and $dp/dx$ is the pressure gradient along the flow direction $x$. In calculating the pressure, we added corrections to the cross-section area by excluding the depletion layer at the boundaries, and $d$ is redefined as the effective thickness where the water-free region (~0.5 nm in total) from the basal plane is excluded as well. We plot $Q$ against $d$ at a constant $dp/dx$ from 20.0 to 100.0 MPa/nm in **Figure 4a**. The results show that the scaling relation between $Q$ and $d$ deviating from cubic ($Q \sim d^3$) towards quartic ($Q \sim d^4$). This can be explained by the reduction of $\eta$ as $d$ increases, which can be evaluated by fitting the velocity profile to the Poiseuille law (**Figure S4**). It should be noticed that this $\eta$-$d$ dependence follow the same trend with other hydroxyl-containing surfaces but contradicts with that observed for water in carbon nanotubes [6, 22-24], which may be correlated to their hydrophobicity.

From the $Q$-$d$ dependence, we quantified the flow enhancement parameter $\varepsilon = \gamma/\gamma_{\text{no-slip}}$, where the hydraulic conductivity $\gamma$ and $\gamma_{\text{no-slip}}$ was calculated from MD simulations, i.e. $\gamma = v/(dp/dx)$, and **Eq. (2)**, respectively. The results shown in **Figure 4b** and **S5** were calculated by using the bulk viscosity of water, which shows that $\varepsilon$ firstly decreases and then increases with $d$. In comparison with bulk water flow that can be described by the Poiseuille law, high values of $\gamma$ at small and large $d$ (< 3 nm), corresponding to the enhancement by interfacial slip and reduction in viscosity, respectively. By adding corrections from viscosity and interfacial slip, we found that the flow enhancement is majorly contributed by the interfacial slippage, while the effect of reduced viscosity becomes important only at large $d$ (**Figure 4**). The enhancement increases with the pressure gradient in the latter mechanism as well. The slip boundary condition could be finally incorporated into the hydrodynamics, yielding a modified prediction for $Q$

$$Q = -(1+6\eta/\beta d)(dp/dx)\, d^3/12\eta, \qquad (3)$$

where $\beta = -(dp/dx)d/2v_s$ and the enhancement by slip is the second term. By using the modified values of $\eta$, **Eq. (3)** could perfectly predict the water flow between GO sheets from the MD simulations (**Figure 4a**), confirming that the essential physics is correctly captured in the model.

## IV. CONCLUSION

In summary, we revealed the breakdown of fast slip flow in chemically functionalized graphene sheets, which could be further weakened by relaxing the nanoconfinement. MD simulation results show that hydrodynamics applies even at scales down to nanometers in this regime, but corrections on the slip boundary condition and viscosity of nanoconfined liquid must included. The contrastive nature from flow between pristine graphene arises from the electrostatic contribution to the interfacial interaction and the modification of H-bond networks. Recent experiments showed that the structure of GO is strongly influenced by the humidity, where over 70% expansion in the stacking direction of GO sheets was observed [11, 25-27]. Thus with a controlled oxidation and reduction processes in fabricating GO, the conclusion from this computational study could thus be verified experimentally.

## ACKNOWLEDGMENTS

This work was supported by the National Natural Science Foundation of China through Grant 11222217, 11002079, Tsinghua University Initiative Scientific Research Program 2011Z02174, and the Tsinghua National Laboratory for Information Science and Technology of China.

**FIGURES AND CAPTIONS**

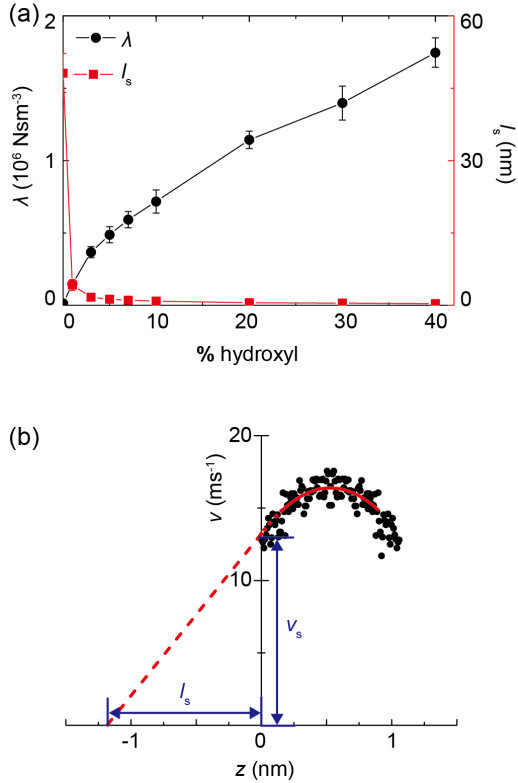

**Figure 1.** (a) Friction coefficient $\lambda$ (black line and circles) and slip length $l_s$ (red line and squares) as functions of the concentration of hydroxyl groups on the functionalized graphene sheet. The error bar of friction coefficients is plotted using their max/min values. (b) One example of the slip length determination in our work. The solid red line is a parabolic fit to the simulation data, the extension of which from $z = 0$ intersects with the x-axis at $-l_s$. The plane $z = 0$ is defined as the average position of the first water layer from GO sheets.

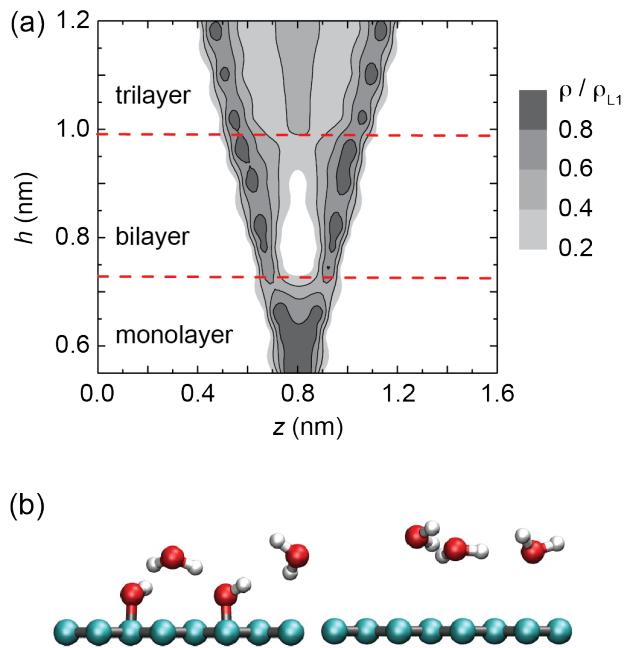

**Figure 2.** (a) Contour plot of the water density distribution $\rho(z, h)$ between GO at $c =$ 10%, with reference to the density of first water layer $\rho_{1st}$. $\rho$ was calculated from MD simulation results for different interlayer distances $h$, measured from basal planes of GO. Gray level in the plot increases with $\rho/\rho_{L1}$. (b) Snapshots of the interfacial structures between water molecules and hydroxyl groups and pristine graphene sheets.

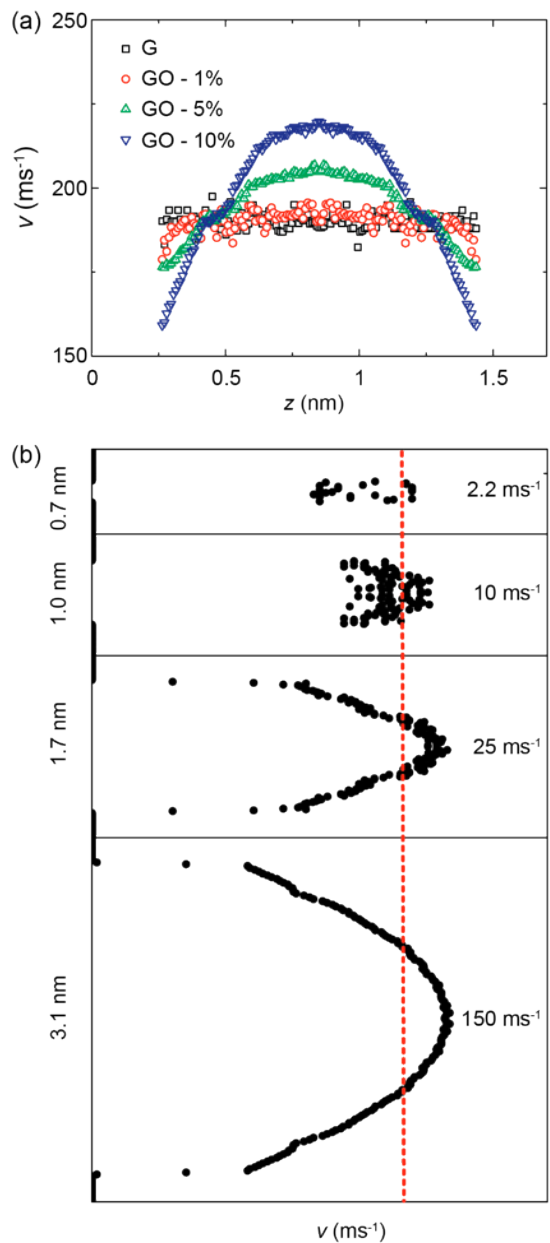

**Figure 3.** (a) Flow velocity profiles of water at different concentration of hydroxyl groups on the graphene sheet. (b) Effects of the interlayer distance $h$ on the flow behavior for $c$ = 10% and $dP/dx$ = 68.5 MPa/nm. The red dash lines indicate a base value as annotated for each subplots.

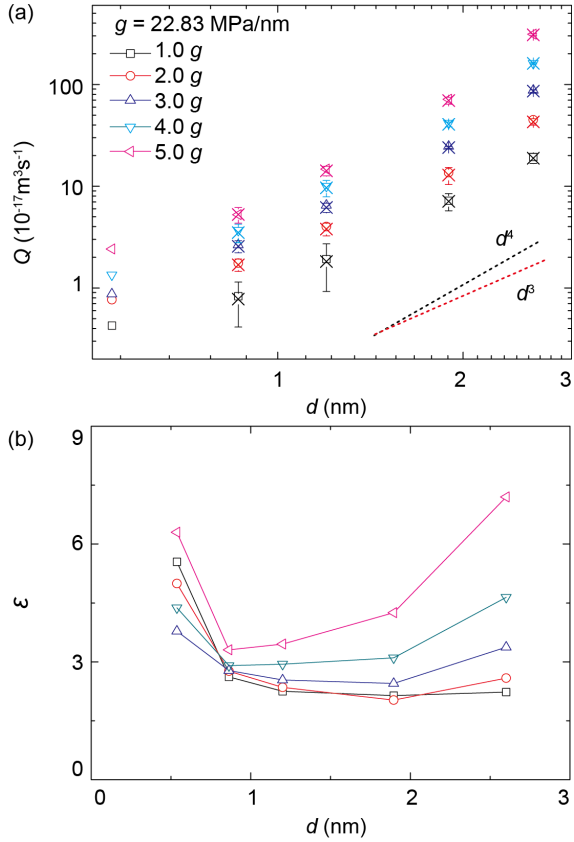

**Figure 4.** (a) Flow rate $Q$ predicted from Poiseuille law with corrections on both the viscosity and boundary slip (**Eq. (3)**) (cross), compared with the simulation results (open symbols) with effective interlayer distance $d$ at hydroxyl groups concentration $c = 10\%$. The pressure gradient is expressed in unit of $g = 22.83$ MPa/nm. Data points for $d = 0.54$ nm is not available for prediction because the absence of parabolic profile (water bilayer forms). (b) Flow enhancement factor $\varepsilon$ as a function of $d$.